%% file: errata.tex
\documentclass{elsart}

\usepackage{graphicx}
\usepackage{subfigure}
\usepackage{amsmath}
\usepackage{xspace}

\newcommand{\PLB}[1]{Phys. Lett. {B #1}}

\def\PRL{Phys. Rev. Lett.}

\newcommand{\mrho}{\mathrm{M}_\rho}

\newcommand{\eepp}{\ensuremath{\rm e^+e^-\to\pi^+\pi^-}\xspace}

\newcommand{\ee}{\ensuremath{\rm e^+e^-}}

\begin{document}

\begin{frontmatter}


\title{Reanalysis of Hadronic Cross Section Measurements at CMD-2}

\author[BINP]{R.R.Akhmetshin},
\author[BINP]{E.V.Anashkin},
\author[JINR]{A.B.Arbuzov},
\author[BINP]{V.Sh.Banzarov},
\author[PITT]{A.Baratt},
\author[BINP]{L.M.Barkov},
\author[BINP,NGU]{A.V.Bogdan},
\author[BINP,NGU]{A.E.Bondar},
\author[BINP]{D.V.Bondarev},
\author[BINP,NGU]{S.I.Eidelman},
\author[BINP,NGU]{D.A.Epifanov},
\author[BINP,NGU]{G.V.Fedotovich},
\author[BINP]{N.I.Gabyshev},
\author[BINP,NGU]{D.A.Gorbachev},
\author[BINP]{A.A.Grebenuk},
\author[BINP,NGU]{D.N.Grigoriev},
\author[YALE]{V.W.Hughes}~\footnote{deceased},
\author[BINP,NGU]{F.V.Ignatov},
\author[BINP]{V.F.Kazanin},
\author[BINP,NGU]{B.I.Khazin},
\author[BINP]{P.P.Krokovny},
\author[JINR]{E.A.Kuraev},
\author[BINP]{L.M.Kurdadze},
\author[BINP,NGU]{A.S.Kuzmin},
\author[BINP,NGU]{Yu.E.Lischenko},
\author[BINP,BU]{I.B.Logashenko},
\author[BINP]{P.A.Lukin},
\author[BINP]{K.Yu.Mikhailov},
\author[BU]{J.P.Miller},
\author[BINP,NGU]{A.I.Milstein},
\author[BINP,NGU]{M.A.Nikulin},
\author[BINP]{A.S.Popov}, 
\author[BU]{B.L.Roberts},
\author[BINP]{N.I.Root},
\author[BINP]{N.M.Ryskulov},
\author[BINP]{Yu.M.Shatunov},
\author[BINP,NGU]{B.A.Shwartz},
\author[BINP]{A.L.Sibidanov},
\author[BINP]{V.A.Sidorov},
\author[BINP]{A.N.Skrinsky},
\author[WEIZ]{V.P.Smakhtin},
\author[BINP,NGU]{E.P.Solodov},
\author[BINP]{P.Yu.Stepanov},
\author[PITT]{J.A.Thompson},
\author[BINP,NGU]{A.S.Zaitsev}
\address[BINP]{Budker Institute of Nuclear Physics, 
Novosibirsk, 630090, Russia}
\address[JINR]{Joint Institute of Nuclear Research,
Dubna, 141980, Russia}
\address[PITT]{University of Pittsburgh, Pittsburgh, PA 15260, USA}
\address[NGU]{Novosibirsk State University, 
Novosibirsk, 630090, Russia}
\address[YALE]{Yale University, New Haven, CT 06511, USA}
\address[BU]{Boston University, Boston, MA 02215, USA}
\address[WEIZ]{Weizmann Institute of Science, Rehovot 76100, Israel}
\vspace{-3mm}
\begin{abstract}
The updated results of the precise measurements
of the processes $\ee \to \rho \to \pi^+\pi^-$, 
$\ee \to \omega \to \pi^+\pi^-\pi^0$
and $\ee \to \phi \to \rm K^0_L K^0_S$ performed by the CMD-2
collaboration are presented. 
The update appeared necessary due 
an overestimate of the integrated luminosity in previous analyses.
 
\end{abstract}
\end{frontmatter}

\section{Introduction}

Precise measurement of the $\ee \to  hadrons$ cross section 
at low energy 
is important for numerous applications in particle physics. The
widely discussed one is the evaluation of the hadronic contribution 
to the muon anomalous magnetic moment. Recent publications hint at
a possible discrepancy between the measurement~\cite{E821} and 
the Standard Model prediction of $a_\mu$~\cite{DEHZ,HMNT}. Analysis 
in Ref.~\cite{DEHZ} also shows inconsistency between the cross 
sections of $\ee \to hadrons$ and the spectral functions of
$\tau \to \nu_\tau$ + hadrons related to the former via conservation
of vector current (CVC).

Since data taking started in 1992, the CMD-2 collaboration
measured various cross sections of $\ee \to hadrons$ in the
c.m.\ energy range 0.36-1.4 GeV and updated parameters of the
$\rho$(770), $\omega$(782) and $\phi$(1020) resonances. All these
results are based on luminosity determined using large angle Bhabha
scattering: $\rm L=N_{\ee}/\tilde{\sigma}_{\ee}$, where $\rm N_{\ee}$ is 
the number of $\ee\to\ee$  detected events and 
$\tilde{\sigma}_{\ee}$ is the
cross section of the process $\ee \to \ee$ in the solid angle of 
the detector with radiative 
corrections taken into account according to~\cite{rcee}. The
radiative corrections include all effects of initial and final state 
radiation and their interference as well as leptonic and hadronic
vacuum polarization.

\begin{figure}
\begin{center}
\includegraphics[width=0.9\textwidth]{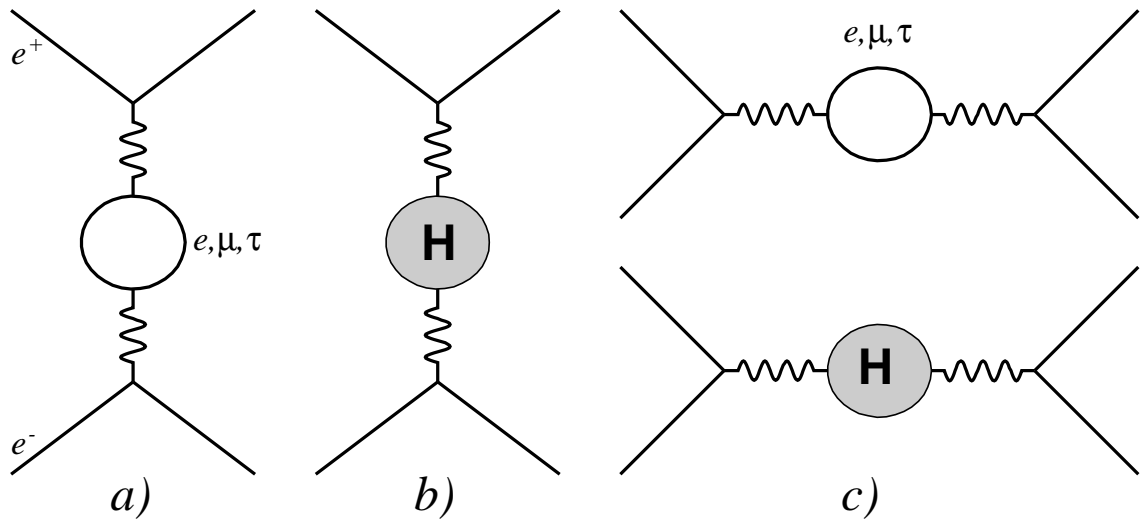}
\end{center}
\caption{\label{loop} Diagrams of the vacuum polarization 
contribution to the $\ee\to\ee$ cross section: a) t-channel (leptons), 
b) t-channel (hadrons), c) s-channel (leptons and hadrons).}
\end{figure}

Recently we found out that the contribution 
from the leptonic loop in the t-channel (Fig.~\ref{loop}a) was
omitted in the computer code for the  
calculation of the radiative corrections to the 
cross section $\tilde{\sigma}_{\ee}$. 
All other loop contributions (Figs.~\ref{loop}b,c) were taken into
account. As a result, the cross-section  $\ee \to \ee$ was
underestimated and the luminosity was overestimated by $2\%
\div 3\%$ depending on energy. 

After fixing the above error,  the total cross section and
various angular and energy distributions calculated in
our code were compared  to the well known 
program BHWIDE, the high precision Monte Carlo generator of the 
Bhabha scattering~\cite{bhwide}. 
Agreement at the level of 0.1\% was found.  Since
the mistake was found in the computer code rather than in the approach
based on Ref.~\cite{rcee}, which was  independently checked
and shown to be valid, 
the systematic error of the calculated $\ee \to \ee$ cross section
is estimated to be 0.2\% as discussed in~\cite{rcee}. 

We have also updated the computer code 
for the calculation of the $\ee \to \mu^+\mu^-$ cross section
fixing a typo in the term related to final state 
radiation (Eq.~(2.23) in Ref.~\cite{rcee}),
which produced an insignificant effect on the results  
of CMD-2 measurements.
A thorough subsequent cross-check of our Monte Carlo generators
for the processes $\ee \to \mu^+\mu^-$ and $\ee \to \pi^+\pi^-$ 
with the independent computer codes~\cite{KKMC,BABAYAGA} showed 
reasonable agreement within the claimed accuracy. 
The details of the comparison will be discussed in a separate paper. 

Most of the cross section measurements performed with CMD-2 
and published
by now~\cite{phif,4pi,5pi,kskl} have systematic uncertainties
significantly larger than 2\% and it is therefore unreasonable to 
correct the results of these papers for the effect mentioned above
before their complete reanalysis 
is done~\footnote{Our recent result on the 
investigation of the process 
$\ee \to \omega\pi^0 \to \pi^0\pi^0\gamma$~\cite{omegapi}
is already based on the corrected luminosity value and 
should not be updated.}.   

Below we present the results of the reanalysis
of our high precision measurements at the $\rho$~\cite{rhoart},
$\omega$~\cite{ompar} and $\phi$~\cite{phipar} resonances.
Following our paper~\cite{rhoart}, throughout this work we present 
two types of the cross section: $\sigma$ and  $\sigma^0$. 
The former quantity, $\sigma$, is the measured or ``dressed''
cross section, which includes both leptonic and hadronic 
vacuum polarization effects
(Fig.~\ref{loop}c) and
should be used in the approximation of the energy dependence
with resonances. The latter, $\sigma^0$, is the ``bare'' cross 
section, in which the leptonic and hadronic vacuum polarization effects 
are removed: 
$\sigma^0=\sigma \cdot |1-\Pi(s)|^2$~\cite{rcee}, to be used in
various applications including calculations of 
the hadronic contribution to the muon anomalous magnetic moment.

\section{Measurement of the pion form factor}

The data analysis in the pion form factor experiment~\cite{rhoart}
was repeated. The corrected luminosity is 317.3 nb$^{-1}$ or 2.4\%
lower than that quoted in the previous publication, corresponding 
to a data sample of 114000  $\pi^+\pi^-$ events. 
The resulting correction to the form factor is slightly larger than 
the luminosity correction discussed
above due to the correlation between the number of $\ee$ and $\pi^+\pi^-$
pairs introduced by the event separation procedure. Table 1 
in~\cite{rhoart} should be replaced by Table~\ref{table:fpi} of this 
work. We remind that the cross section   $\sigma^0_{\pi\pi(\gamma)}$
shown in the last column of the Table is the ``bare'' cross section
which also includes the effect of final state radiation. The
systematic error of the cross section is estimated to be 0.6\%,
the same as in the original publication~\cite{rhoart}. 

The $\rho$-meson parameters listed in the Conclusion of~\cite{rhoart},
should be replaced with the following values obtained with the
Gounaris-Sakurai parameterization of the reanalysed data:
\begin{equation*}
\begin{array}{ll}
\mrho  & = (775.65 \pm 0.64 \pm 0.50)~ \text{MeV},
\vphantom{\Bigl( \Bigr)} \\
\Gamma_\rho  & = (143.85\pm 1.33 \pm 0.80)~ \text{MeV},
\vphantom{\Bigl( \Bigr)} \\
\Gamma(\rho\rightarrow e^+e^-)  & = (7.06\pm 0.11 \pm 0.05)~ \text{keV},
\vphantom{\Bigl( \Bigr)} \\
\mathcal{B}(\omega\rightarrow\pi^+\pi^-) & = ( 1.30\pm 0.24 \pm 0.05) \% ,
\vphantom{\Bigl( \Bigr)} \\
\mathrm{arg} \, \delta & = 13.3^{\circ} \pm 3.7^{\circ} \pm 0.2^{\circ}. 
\vphantom{\Bigl( \Bigr)} 
\end{array}
\end{equation*}
Compared to the previous analysis~\cite{rhoart}, the mass and 
total width become smaller by 0.54$\sigma$ and 0.39$\sigma$, respectively.
The branching fraction $\mathcal{B}(\omega\rightarrow\pi^+\pi^-)$ and the
quantity  $\mathrm{arg} \, \delta$ vary only slightly. 
Finally, the leptonic width becomes larger by 2.9\% (1.7
standard deviations). 

The hadronic contribution to the muon anomalous magnetic moment 
from the $\pi^+\pi^-$ channel in the energy range covered by present 
analysis is estimated to be $(378.6 \pm 2.7 \pm 2.3) \times 10^{-10}$ 
or $10.5 \times 10^{-10}$ higher than in our previous estimate.

\begin{table*}
\begin{center}
\caption{\label{table:fpi} The measured value of the pion form
factor and ``bare'' cross section $\eepp(\gamma)$. 
Only statistical errors are shown. The systematic error is
estimated to be 0.6\%.}
\begin{tabular}[t]{ccc@{\hspace{10mm}}ccc}
\hline 
E$_{\rm c.m.}$, MeV & $|\rm F_\pi|^2$ & $\sigma^0_{\pi\pi(\gamma)}$, nb &
E$_{\rm c.m.}$, MeV & $|\rm F_\pi|^2$ & $\sigma^0_{\pi\pi(\gamma)}$, nb \\
\hline 
\input{errata.tab}
\hline
\end{tabular}
\end{center}
\end{table*}

\vspace{5mm}

\input{omnew.tex}

\input{phinew.tex}

\section{Conclusion}

We performed a reanalysis of the high precision measurements
of the processes $\ee\to \rho \to \pi^+\pi^-$~\cite{rhoart}, 
$\ee \to \omega \to \pi^+\pi^-\pi^0$~\cite{ompar}
and $\ee \to \phi \to \rm K^0_L K^0_S$~\cite{phipar} at the CMD-2
detector. The corrected values of the $\rho, \omega$ and $\phi$
meson parameters are presented together with the detailed tables of the 
corresponding hadronic cross sections. 



{\large \textbf Acknowledgements}

The authors are grateful to M.N.~Achasov, M.~Davier, V.P.~Druzhinin, 
V.S.~Fadin, S.~Jadach, F.~Jegerlehner and W.~Placzek
for numerous valuable discussions.

This work is supported in part by grants DOE DEFG0291ER40646, 
INTAS 96-0624, NSF PHY-9722600, NSF PHY-0100468, 
RFBR-98-02-1117851 and RFBR-03-02-16843.

\end{document}

%% file: errata.tab
$610.50$ & $ 8.18 \pm 1.15$ & $ 335.2 \pm 47.2$ & $784.24$ & $35.24 \pm 1.00$ & $ 959.8 \pm 27.3$ \\ 
$620.50$ & $ 9.95 \pm 0.74$ & $ 400.2 \pm 29.7$ & $786.04$ & $30.92 \pm 1.10$ & $ 835.0 \pm 29.7$ \\ 
$630.50$ & $11.12 \pm 0.74$ & $ 438.7 \pm 29.1$ & $790.10$ & $33.03 \pm 1.14$ & $ 890.0 \pm 30.8$ \\ 
$640.51$ & $11.30 \pm 0.73$ & $ 437.3 \pm 28.3$ & $794.14$ & $30.90 \pm 0.87$ & $ 830.1 \pm 23.4$ \\ 
$650.49$ & $13.02 \pm 0.85$ & $ 493.9 \pm 32.4$ & $800.02$ & $30.25 \pm 0.73$ & $ 806.2 \pm 19.6$ \\ 
$660.50$ & $13.77 \pm 0.77$ & $ 512.2 \pm 28.7$ & $810.14$ & $26.33 \pm 0.56$ & $ 689.7 \pm 14.8$ \\ 
$670.50$ & $15.79 \pm 0.84$ & $ 576.0 \pm 30.7$ & $820.02$ & $24.78 \pm 0.79$ & $ 637.3 \pm 20.4$ \\ 
$680.59$ & $20.02 \pm 0.93$ & $ 716.0 \pm 33.1$ & $829.97$ & $21.15 \pm 0.76$ & $ 533.9 \pm 19.1$ \\ 
$690.43$ & $21.03 \pm 0.74$ & $ 737.4 \pm 26.1$ & $839.10$ & $17.29 \pm 0.72$ & $ 429.2 \pm 17.9$ \\ 
$700.52$ & $24.46 \pm 0.64$ & $ 840.2 \pm 22.1$ & $849.24$ & $14.57 \pm 0.72$ & $ 355.0 \pm 17.5$ \\ 
$710.47$ & $28.37 \pm 1.01$ & $ 954.6 \pm 34.0$ & $859.60$ & $14.91 \pm 0.70$ & $ 356.5 \pm 16.7$ \\ 
$720.25$ & $32.74 \pm 0.85$ & $1078.8 \pm 27.9$ & $869.50$ & $11.41 \pm 0.47$ & $ 268.0 \pm 11.0$ \\ 
$730.24$ & $35.29 \pm 1.21$ & $1137.1 \pm 39.0$ & $879.84$ & $10.47 \pm 0.81$ & $ 241.4 \pm 18.7$ \\ 
$740.20$ & $38.81 \pm 1.20$ & $1222.2 \pm 37.9$ & $889.72$ & $ 8.69 \pm 0.35$ & $ 197.0 \pm  7.9$ \\ 
$750.28$ & $43.52 \pm 1.17$ & $1338.4 \pm 35.9$ & $900.04$ & $ 8.04 \pm 0.30$ & $ 178.9 \pm  6.7$ \\ 
$760.18$ & $44.71 \pm 1.19$ & $1343.6 \pm 35.7$ & $910.02$ & $ 7.05 \pm 0.32$ & $ 154.1 \pm  7.0$ \\ 
$764.17$ & $44.48 \pm 1.05$ & $1325.6 \pm 31.4$ & $919.56$ & $ 6.23 \pm 0.31$ & $ 134.0 \pm  6.7$ \\ 
$770.11$ & $45.48 \pm 1.17$ & $1340.3 \pm 34.5$ & $930.11$ & $ 5.91 \pm 0.37$ & $ 124.9 \pm  7.9$ \\ 
$774.38$ & $44.29 \pm 1.17$ & $1297.5 \pm 34.4$ & $942.19$ & $ 5.35 \pm 0.25$ & $ 110.8 \pm  5.3$ \\ 
$778.17$ & $46.68 \pm 1.33$ & $1358.0 \pm 38.7$ & $951.84$ & $ 4.72 \pm 0.24$ & $  96.2 \pm  4.9$ \\ 
$780.17$ & $44.04 \pm 1.22$ & $1265.5 \pm 35.1$ & $961.52$ & $ 4.44 \pm 0.23$ & $  89.1 \pm  4.7$ \\ 
$782.23$ & $37.55 \pm 0.68$ & $1049.3 \pm 18.9$ & \\ 

%% file: omnew.tex

\section{Measurement of the $\omega(782)$ meson 
parameters in the $\omega \to \pi^+\pi^-\pi^0$ mode}

In the reanalysis of our $\omega$ meson experiment,
the selection procedure described in Ref.~\cite{ompar}
can be left unchanged. Although small, the background
cross section is not negligible, therefore
a refit of the data is necessary. 

The corrected integrated luminosity is 119.6 nb$^{-1}$, i.e. 2.4\%
smaller than before. The results of the fit based on a data sample
of about 11200 events are presented below:

\begin{equation*}
\begin{array}{ll}
\rm M_\omega  & = (782.68 \pm 0.09 \pm 0.04)~ \text{MeV},
\vphantom{\Bigl( \Bigr)} \\
\Gamma_\omega  & = (8.68\pm 0.23 \pm 0.10)~ \text{MeV},
\vphantom{\Bigl( \Bigr)} \\
\sigma_0  & = (1495.6\pm 25.5 \pm 19.4)~ \text{nb},
\vphantom{\Bigl( \Bigr)} \\
\sigma_{bg}  & = (12.2\pm 4.5)~ \text{nb}.
\vphantom{\Bigl( \Bigr)} 
\end{array}
\end{equation*} 

Comparison with the previous publication shows that the values of 
the total width
and background cross section remain the same while the value of the
mass becomes 0.03~MeV smaller. The value of the cross section at the
peak is 2.6\% higher than previously.

From the value of the cross section at the peak, $\sigma_0$,
one can calculate the following product of the branching ratios:
\begin{equation*} 
\mathcal{B}(\omega \to \ee)\mathcal{B}(\omega \to \pi^+\pi^-\pi^0) = 
(6.24\pm 0.11 \pm 0.08) \times 10^{-5} .
\end{equation*}

Finally, in Table~\ref{tabom} we present the measured
and ``bare'' cross sections as a function of c.m. energy. 
The systematic error of the cross section is estimated to be 1.3\%,
the same as in the original publication~\cite{ompar}.
 
\vspace{4mm}

\begin{table}[h]
\begin{center}
\caption{The measured
and  ``bare'' cross section of the process 
$\omega \to \pi^+\pi^-\pi^0$. Only statistical
errors are shown. The systematic error is
estimated to be 1.3\%.}
\label{tabom}
\begin{tabular}{ccc}
\hline
E$_{\rm c.m.}$, MeV & $\sigma$, nb & $\sigma^0$, nb \\
\hline
760.18 & $69.0 \pm 11.0$  &  $68.0 \pm 10.8$ \\
764.17 & $71.0 \pm 8.0$  &  $70.0 \pm 7.9$ \\
770.11 & $179.0 \pm 15.0$  &   $176.6 \pm 14.8$ \\
774.38 & $279.0 \pm 22.0$  &  $275.9 \pm 21.8$ \\
778.17 & $790.0 \pm 41.0$  &  $781.8 \pm 40.6$ \\
780.17 & $1193.0 \pm 51.0$ &  $1149.1 \pm 50.5$ \\
782.23 & $1490.0 \pm 31.0$ &  $1427.8 \pm 29.7$ \\
784.24 & $1338.0 \pm 45.0$ &  $1254.8 \pm 42.2$ \\
786.04 & $903.0 \pm 44.0$  & $842.8 \pm 41.1$ \\
790.09 & $417.0 \pm 19.0$  &  $391.4 \pm 17.8$ \\
794.14 & $197.0 \pm 11.0$  &  $185.9 \pm 10.4$ \\
800.00 & $127.0 \pm 8.0$  &  $120.3 \pm 7.6$ \\
810.14 & $56.0 \pm 4.0$  &  $53.2 \pm 3.8$ \\
\hline
\end{tabular}
\end{center}
\end{table}

%% file: phinew.tex

\section{Measurement of the $\phi$(1020) meson 
parameters in the $\phi \to \rm K^0_LK^0_S$ mode}

The changes in the results of our  $\phi$(1020) meson study in  
the $\phi\to \rm K^0_LK^0_S$ mode~\cite{phipar} are straightforward.
The corrected integrated luminosity is 1924 nb$^{-1}$ or 2.7\%
smaller than previously. The corresponding data sample contains
 2.72$\times$10$^5$ $\rm K^0_LK^0_S$ events obtained in the analysis
of four independent scans.

 The reanalysis showed that the
peak cross section, $\sigma_0(\phi \to \rm K^0_LK^0_S)$,
as well as the values of the product of the branching fractions
$\mathcal{B}_{\rm e^+e^-}\mathcal{B}_{\rm K^0_LK^0_S}$ become  
higher by the same amount of 2.7\%. 
The values of the $\phi$ mass and total width remain unchanged.
The new values of  
$\phi$(1020) parameters in four scans of our experiment
are presented in Table~\ref{tab:1}.

\vspace{4mm}

\begin{table}[h]
\begin{center}
\caption{ $\phi$ meson parameters obtained in this analysis}
\label{tab:1}
\vspace*{2mm}
\begin{tabular}{|c|c|c|c|c|}
\hline
Scan & $\sigma_0$, nb     & $\rm m_{\phi}$, MeV/c$^2$ & 
$\mathcal{B}_{\rm e^+e^-}\mathcal{B}_{\rm K^0_LK^0_S}, 10^{-4}$  \\
\hline
 1   & $1402\pm 14\pm 24$ & $1019.506\pm 0.030\pm 0.020$ & 
$0.993\pm 0.010\pm 0.016$\\
 2   & $1378\pm 13\pm 24$ & $1019.512\pm 0.023\pm 0.045$ &
$0.976\pm 0.009\pm 0.016$\\
 3   & $1434\pm 12\pm 25$  & $1019.363\pm 0.017\pm 0.080$ & 
$1.016\pm 0.008\pm 0.017$\\
 4   & $1431\pm 12\pm 25$ & $1019.316\pm 0.021\pm 0.122$ &
$1.013\pm 0.008\pm 0.017$\\
\hline
Average & $1413\pm 6\pm 24$ & $1019.483\pm 0.011\pm 0.025$ &
$1.001\pm 0.004\pm 0.017$\\
\hline
\end{tabular}
\end{center}
\end{table}

The value of the $\phi$ meson total width is the same as before:
$$
	\Gamma_{\phi} = (4.280 \pm 0.033 \pm 0.025)\ \rm MeV.
$$


The measured and ``bare'' cross sections of the process 
$\phi\to \rm K^0_LK^0_S$ as a function of c.m. energy are presented 
in Table~\ref{f:allscans}. The systematic error of the cross section
is estimated to be 1.7\%, the same as in the original 
publication~\cite{phipar}.

\begin{table}[h]
\begin{center}
\caption{The measured and ``bare''
cross section of the process $\phi\to \rm K^0_LK^0_S$ for all scans.
Only statistical errors are shown. The systematic error is estimated
to be 1.7\% for all four scans.}

\label{f:allscans}
\vspace*{5mm}
{\small
\hspace*{-10mm}\begin{tabular}{ccc|ccc}
\hline
E$_{c.m.}$, MeV & $\sigma$, nb & $\sigma^0$, nb &
E$_{c.m.}$, MeV & $\sigma$, nb & $\sigma^0$, nb \\
\hline
\multicolumn{3}{c|}{1 Scan} & \multicolumn{3}{c}{2 Scan}\\
\hline
1010.27$\pm$0.03 &  42.21$\pm$5.16&   42.89$\pm$5.24  & 1004.25$\pm$0.17 & 18.51$\pm$ 9.85&18.40$\pm$9.79    \\  
1017.09$\pm$0.02 & 602.85$\pm$14.91& 658.31$\pm$16.28 & 1010.86$\pm$0.13  &  52.96$\pm$ 7.53 & 53.97$\pm$ 7.67   \\  
1018.14$\pm$0.02 & 999.68$\pm$34.84&1069.66$\pm$37.28 & 1016.37$\pm$0.08 & 399.54$\pm$35.28 & 433.50$\pm$ 38.28    \\
1018.96$\pm$0.02 &1278.75$\pm$32.27&1277.47$\pm$32.24 & 1017.19$\pm$0.08 & 600.22$\pm$45.78 & 655.44$\pm$49.99    \\
1019.21$\pm$0.02 &1328.94$\pm$38.80&1291.73$\pm$37.71 & 1018.06$\pm$0.08 & 930.66$\pm$51.35 & 999.53$\pm$55.15   \\
1019.99$\pm$0.02 &1325.08$\pm$28.63&1189.92$\pm$25.71 & 1019.00$\pm$0.08 &1329.00$\pm$25.08&1322.36$\pm$24.95    \\ 
1020.13$\pm$0.02 &1342.71$\pm$41.89&1193.67$\pm$37.24 & 1020.00$\pm$0.08 &1282.51$\pm$50.32&1150.41$\pm$45.14    \\ 
1021.85$\pm$0.02 & 622.82$\pm$33.88& 536.87$\pm$29.20 & 1020.96$\pm$0.08 & 941.38$\pm$46.99 & 811.47$\pm$40.51    \\   
1023.97$\pm$0.02 & 292.26$\pm$14.91& 258.07$\pm$13.17 & 1021.88$\pm$0.09 & 620.70$\pm$40.29 & 535.04$\pm$34.73   \\
\multicolumn{3}{c|}{} & 1027.70$\pm$0.11 & 126.74$\pm$10.35&115.46$\pm$9.43\\
\multicolumn{3}{c|}{} & 1033.63$\pm$0.17 &  66.33$\pm$8.57 & 61.69$\pm$7.97\\   
\multicolumn{3}{c|}{} & 1039.48$\pm$0.17 &  37.92$\pm$ 6.23& 35.61$\pm$5.85\\    
\hline
\hline
\multicolumn{3}{c|}{3 Scan}& \multicolumn{3}{c}{4 Scan} \\
\hline
1004.64$\pm$0.17 &   13.58$\pm$ 4.59 &  13.51$\pm$4.57 &1004.19$\pm$0.17 &   12.39$\pm$1.77 & 12.32$\pm$1.76   \\   
1011.30$\pm$0.09 &   52.97$\pm$ 3.48 &  54.14$\pm$3.56 &1011.30$\pm$0.10 &   56.62$\pm$6.87 & 57.87$\pm$7.02    \\   
1015.99$\pm$0.08 &  350.79$\pm$28.31 & 378.50$\pm$30.55&1015.91$\pm$0.08 &  343.95$\pm$26.62 & 370.78$\pm$28.70    \\   
1016.93$\pm$0.08 &  560.58$\pm$42.85 & 611.59$\pm$46.75&1016.94$\pm$0.08 &  601.65$\pm$45.64 & 656.40$\pm$49.79    \\   
1017.91$\pm$0.08 &  931.61$\pm$49.23 &1006.14$\pm$53.17&1017.92$\pm$0.08 &  998.50$\pm$51.38 & 1078.38$\pm$55.49    \\   
1019.04$\pm$0.07 & 1354.29$\pm$25.21 &1342.10$\pm$24.98&1018.76$\pm$0.08 & 1317.09$\pm$23.21 &1344.75$\pm$23.70    \\   
1019.95$\pm$0.07 & 1251.84$\pm$49.67 &1127.91$\pm$44.75&1019.68$\pm$0.07 & 1321.09$\pm$45.42 & 1219.37$\pm$41.92   \\   
1020.86$\pm$0.08 &  891.48$\pm$45.54 & 770.24$\pm$39.35&1020.68$\pm$0.08 &  999.30$\pm$49.45 & 866.39$\pm$42.87    \\   
1021.74$\pm$0.08 &  606.96$\pm$37.01 & 522.59$\pm$31.87&1021.60$\pm$0.08 &  648.54$\pm$36.18 & 558.39$\pm$31.15    \\   
1022.67$\pm$0.09 &  419.31$\pm$30.91 & 364.38$\pm$26.86&1022.59$\pm$0.08 &  428.05$\pm$27.35 & 371.98$\pm$23.77    \\   
1028.36$\pm$0.12 &  102.38$\pm$ 9.75 &  93.58$\pm$8.91 &1028.41$\pm$0.10 &  102.57$\pm$8.42 & 93.75$\pm$7.70    \\   
1034.06$\pm$0.17 &   54.04$\pm$ 7.78 & 50.31$\pm$7.24 &
\multicolumn{3}{c}{}\\
\hline
\hline
\end{tabular}
}
\end{center}
\end{table}
